# Development of a method for measuring blood coagulation using superparamagnetic iron oxide nanoparticles and an alternating magnetic field

Kenya Murase[*]

Department of Medical physics and Engineering, Division of Medical Technology and Science, Course of Health Science, Graduate School of Medicine, Osaka University, 1-7 Yamadaoka, Suita, Osaka 565-0871, Japan
*Corresponding author, email: murase@sahs.med.osaka-u.ac.jp

**Abstract**
We developed a method for measuring blood coagulation using superparamagnetic iron oxide nanoparticles (SPIONs) and an alternating magnetic field (AMF). The 3rd and 5th harmonic signals from SPIONs mixed with blood induced by AMF were detected using a gradiometer coil. Blood coagulation was induced artificially by adding $CaCl_2$ solution to whole blood of sheep at various temperatures and hematocrits. We calculated the coagulation rate ($k$) and normalized signal intensity at infinite time ($S_{inf}$) by fitting the time course of the normalized 3rd harmonic signal to $S(t) = (1 - S_{inf})e^{-kt} + S_{inf}$. The $k$ values increased significantly with increasing temperature and decreased significantly with increasing hematocrit. The $S_{inf}$ values decreased significantly with increasing temperature and tended to increase with increasing hematocrit. Blood anticoagulation was induced by adding heparin to the whole blood sampled from mice. There were significant differences in both the 3rd and 5th harmonic signals between groups with and without heparin at 25 min or more after adding heparin. We also calculated the 3rd and 5th harmonic signals for viscosities ranging from 0.001 to 1 kg/m/s, with an assumption that the magnetization and particle size distribution of SPIONs obey the Langevin theory of paramagnetism and log-normal distribution, respectively. The 3rd and 5th harmonic signals increased slowly with increasing viscosity and had peaks at approximately 0.015 and 0.025 kg/m/s, respectively. After these peaks, they decreased monotonically with increasing viscosity. These results confirm the rationale of our method. In conclusion, our method will be useful for measuring blood coagulation and anticoagulation and for studying their processes.

## 1. Introduction

Blood coagulation is a complex physiological process that normally is vital to the preservation of life. Although blood clotting is essential to stop bleeding, blood clots that impede the flow of blood *in vivo* are responsible for most heart attacks and strokes and complicate other pathological conditions, including many types of cancer and peripheral vascular disease [1].

To assess blood coagulation and its inhibition is crucial for clinical diagnosis and for determination of therapeutic strategies. A common method to assess the process of blood coagulation involves adding blood to three or four test tubes and then tilting the tubes at 30-s intervals until the blood can no longer flow [2]. The current assays frequently used to assess blood coagulation properties are based on mechanical impedance, electromagnetism, rheometry, ultrasound, and photometry [2]. Recently, blood coagulation testing has also been implemented using microfluidic systems [3]. In these systems, blood coagulation characterization was performed based on an increase in blood viscosity [4]. The ability to detect viscosity changes that accompany biological events such as blood coagulation and anticoagulation makes these systems a viable method for monitoring such processes.

Recently, a new imaging method called magnetic particle imaging (MPI) has been introduced [5]. MPI allows imaging of the spatial distribution of magnetic nanoparticles (MNPs) such as superparamagnetic iron oxide nanoparticles (SPIONs) with high sensitivity, high spatial resolution, and high imaging speed. MPI uses the nonlinear response of MNPs to detect their presence in an alternating magnetic field (AMF) called the drive magnetic field. Spatial encoding is realized by saturating the MNPs almost everywhere except in the vicinity of a special point called the field-free point (FFP) using a static magnetic field (selection magnetic field) [5].

Due to the nonlinear response of MNPs to an applied AMF, the signal induced by MNPs in a receive coil contains the excitation frequency as well as the harmonics of this frequency. These harmonics are strongly related to the magnetic relaxation [6, 7]. The dominant relaxation mechanisms are Néel relaxation for smaller MNPs and Brownian relaxation for larger MNPs. In the Néel relaxation mechanism, internal reorientation of the magnetic moment of MNPs occurs, whereas physical rotation of MNPs occurs in the Brownian relaxation mechanism so it depends on a wide array of environmental factors including viscosity and chemical binding [6, 7]. Brownian relaxation is sufficiently sensitive to have been used to measure temperature [8], viscosity [9], and chemical binding [10]. If we could integrate a method for measuring such environmental factors into the above MPI system, it would provide a new diagnostic and/or therapeutic strategy based on imaging in clinical settings.

The purpose of this study was to develop a method for measuring blood coagulation and its inhibition using SPIONs and AMF. We also conducted a theoretical analysis of whether this method has the ability to detect viscosity changes that accompany blood coagulation and its inhibition.

## 2. Materials and methods

### 2.1. Experimental studies

### 2.1.1. Apparatus

Figure 1 illustrates our apparatus for measuring blood coagulation. The AMF was generated using an excitation (solenoid) coil (130 mm in length, 60 mm in inner diameter, and 80 mm in outer diameter), whose AC power was supplied by a programmable power supply (EC1000S, NF Co., Kanagawa, Japan). The AC power was controlled by use of a cosine wave with a frequency of 400 Hz, which was generated using a digital function generator (DF1906, NF Co., Kanagawa, Japan). The peak-to-peak strength of the AMF was taken as 20 mT in this study. The signal generated by MNPs was received by a gradiometer coil (60 mm in length, 25 mm in inner diameter, and 30 mm in outer diameter) consisting of two solenoid coils placed in series and connected differentially with opposite winding directions [11], and the 3rd and 5th harmonic signals were extracted using a preamplifier (T-AMP03HC, Turtle Industry Co., Ltd., Ibaragi, Japan) and a lock-in amplifier (LI5640, NF Co., Kanagawa, Japan). The output of the lock-in amplifier was converted to digital data by a personal computer (PC) with a data acquisition card (C-Logger, Contec Co., Ltd., Osaka, Japan). The sampling duration and total sampling time were taken as 100 μs and 1 s, respectively.

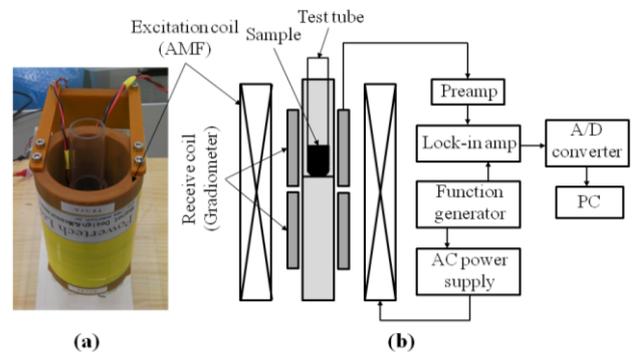

**Figure 1:** Photograph (a) and diagram (b) of our apparatus for measuring blood coagulation. The alternating magnetic field (AMF) was generated using an excitation (solenoid) coil whose AC power was supplied by a programmable power supply. The signals generated by superparamagnetic iron oxide nanoparticles (SPIONs) were received by a gradiometer coil, and the 3rd and 5th harmonic signals were extracted using a preamplifier and a lock-in amplifier.

### 2.1.2. Magnetic nanoparticles

We used a magnetic fluid (M-300) (Sigma High Chemical Co., Ltd., Kanagawa, Japan) as the MNPs (SPIONs). M-300 consists of magnetite ($Fe_3O_4$) [7] and 5% sodium tetradecene sulfonate as a surfactant. Figure 2(a) shows a photograph of M-300 taken by a transmission electron microscope (TEM) (Hitachi H-7650, Hitachi Ltd., Tokyo, Japan), and figure 2(b) shows the magnetization curve of M-300, which was measured at room temperature by use of a vibrating sample magnetometer (Model BHV-525, Riken Denshi Co. Ltd., Tokyo, Japan).

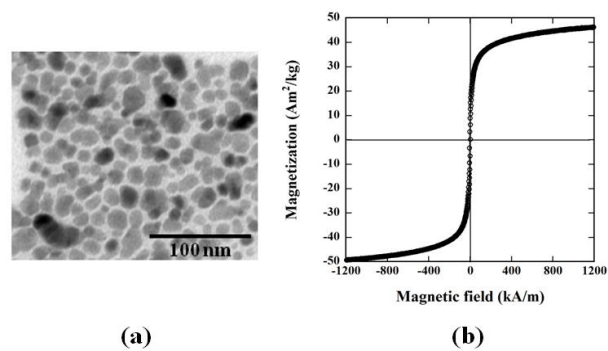

**Figure 2:** (a) Photograph of a transmission electron microscope (TEM) image of the SPIONs used in this study (M-300). The calibration bar (100 nm) is also shown. (b) Magnetization curve of M-300 as a function of magnetic field, which was measured using a vibrating sample magnetometer.

We measured the particle size distribution of M-300 from the TEM image (figure 2(a)) by putting a cursor on both sides of each particle on the TEM image and measuring the distance between them. We obtained the particle size of M-300 as 17.1±4.94 nm (mean±standard deviation (SD)) by fitting the data to a log-normal distribution [7]. We also obtained the mean and SD of the log-normal distribution as 2.80 and 0.282, respectively [7].

### 2.1.3. Blood coagulation

Whole sheep blood was purchased from Kohjin Bio Co., Ltd. (Saitama, Japan). It was preserved in Alsever's solution to prevent coagulation and the hematocrit was 45.0%. First, the whole blood (400 μL) was mixed with M-300 (5 μL) and then blood coagulation was induced artificially by adding 0.5 M $CaCl_2$ solution (40 μL) to the whole blood including M-300.

The 3rd and 5th harmonic signals were measured using our apparatus illustrated in figure 1 immediately before, and 5, 10, 30, 60, 90, and 120 min after adding $CaCl_2$ solution. Unless otherwise specifically stated, the measurements were performed at room temperature (27 °C). To investigate the dependency of blood coagulation on temperature, the measurements were also performed while keeping the blood sample at 37 °C using warmed water.

When we investigated the dependency of blood coagulation on hematocrit, we performed the following procedure prior to the measurement of blood coagulation. First, the whole blood was centrifuged at 6400 rpm for 12 min to separate the blood into the composition of packed erythrocytes and the plasma. The red blood cells were then rinsed twice using a buffered saline solution. Finally, the whole blood sample was mixed as erythrocytes and plasma at a certain volume ratio. This ratio is represented by hematocrit.

### 2.1.4. Quantification of coagulation rate

We calculated the coagulation rate ($k$) and the normalized signal intensity at infinite time ($S_{inf}$) by fitting the time course of the 3rd harmonic signal normalized by that immediately before adding $CaCl_2$ solution ($S(t)$) to the following equation:

$$S(t) = (1 - S_{inf})e^{-kt} + S_{inf} \quad (1)$$

In this study, we used the Simplex method [12] for curve fitting.

### 2.1.5. Blood anticoagulation

Whole blood was obtained by cardiac puncture from 7 male ICR mice aged 8 weeks. The hematocrit was 46.5±1.1% (mean±SD). Heparin sodium ($10^6$ U/L) (Mochida Pharmaceutical Co., Ltd., Tokyo, Japan) was used as an anticoagulant. First, heparin sodium (10 μL) and M-300 (5 μL) were mixed in a test tube and then the whole blood sampled from mice (400 μL) was added to the test tube to investigate the effect of heparin on blood anticoagulation, at 5 min after the completion of blood sampling. The 3rd and 5th harmonic signals were measured in the same manner as mentioned above at 5, 10, 30, 60, and 120 min after the completion of blood sampling, *i.e.*, immediately before, and then 5, 25, 55, and 115 min after adding the whole blood to the test tube including heparin sodium and M-300. The above procedures were performed at room temperature (27 °C). These animal experiments were approved by the animal ethics committee at Osaka University School of Medicine.

### 2.1.6. Statistical analysis

When investigating the effects of temperature and hematocrit on blood coagulation, the differences of the 3rd and 5th harmonic signals among groups were analyzed by one-way analysis of variance (ANOVA). The differences in the $k$ and $S_{inf}$ values among groups with different hematocrits were also analyzed by ANOVA. Statistical significance was determined by Holm's multiple comparison test [13]. When investigating the effect of heparin sodium on blood anticoagulation, the differences of the 3rd and 5th harmonic signals between groups with and without heparin sodium were analyzed by unpaired t-test. The differences in the $k$ and $S_{inf}$ values between groups with different temperatures were also analyzed by unpaired t-test. A *P* value less than 0.05 was considered statistically significant.

## 2.2. Simulation studies

We performed simulation studies according to the following procedure to confirm the rationale of our method.

### 2.2.1. Signal detection

Assuming a single receive coil with sensitivity ($\sigma_r(\boldsymbol{r})$) at spatial position $\boldsymbol{r}$, the changing magnetization of MNPs induces a voltage according to Faraday's law, which is given by [14]

$$v(t) = -\mu_0 \frac{d}{dt} \int_\Omega \sigma_r(\boldsymbol{r}) M(\boldsymbol{r}, t) d\boldsymbol{r} \quad (2)$$

where $\Omega$ denotes the volume containing MNPs, $M(\boldsymbol{r}, t)$ is the magnetization generated by MNPs at position $r$ and time $t$, and $\mu_0$ is the magnetic permeability of the vacuum. The receive coil sensitivity ($\sigma_r(r)$) derives from the magnetic field that the coil would produce if driven with a unit current [14].

In the following, the receive coil sensitivity was assumed to be constant and uniform over the volume of interest and to be denoted by $\sigma_0$. Thus, $v(t)$ given by equation (2) is reduced to

$$v(t) = -\mu_0 \sigma_0 \frac{d}{dt} \int_\Omega M(r,t) dr \quad (3)$$

Neglecting constant factors, we introduce the notation $s(t)$ for the signal generated by MNPs. If we denote the magnetization after volume integral as $M(t)$, the signal can then be given by

$$s(t) = -\frac{dM(t)}{dt} \quad (4)$$

The Fourier transformation of both sides of equation (4) yields

$$\mathcal{F}[s(t)] = \mathcal{F}\left[-\frac{dM(t)}{dt}\right] = -2\pi f \mathcal{F}[M(t)] \quad (5)$$

where $\mathcal{F}[\cdot]$ and $f$ denote the Fourier transform and the frequency, respectively. The spectrum was calculated by taking the absolute value of equation (5).

### 2.2.2. Langevin function

It is convenient to express $M(t)$ in response to an applied AMF in terms of the complex susceptibility ($\chi$) as [15]

$$M(t) = \text{Re}[\chi H(t)] \quad (6)$$

where $H(t)$ is the AMF at time $t$ and Re denotes the real part of a complex number. In this study, we assumed that $H(t)$ in equation (6) is given by

$$H(t) = H_{ac}\cos(2\pi f_{ac} t) = \text{Re}(H_{ac} e^{j2\pi f_{ac} t}) \quad (7)$$

where $H_{ac}$ and $f_{ac}$ denote the amplitude and frequency of the AMF, respectively, and $j = \sqrt{-1}$. $\chi$ in equation (6) is given by [15]

$$\chi = \chi' - j\chi'' \quad (8)$$

where $\chi'$ and $\chi''$ are the in-phase and out-of-phase components of $\chi$, respectively, and are given by [15]

$$\chi' = \frac{\chi_0}{1+(2\pi f_{ac} \tau)^2} \text{ and } \chi'' = \frac{2\pi f_{ac} \tau \chi_0}{1+(2\pi f_{ac} \tau)^2} \quad (9)$$

In equation (9), $\chi_0$ is the equilibrium susceptibility and $\tau$ is the effective relaxation time given by

$$\frac{1}{\tau} = \frac{1}{\tau_N} + \frac{1}{\tau_B} \quad (10)$$

where $\tau_N$ and $\tau_B$ are the Néel relaxation time and Brownian relaxation time, respectively. $\tau_N$ and $\tau_B$ are given by the following relationships [15]:

$$\tau_N = \tau_0 \frac{\sqrt{\pi} e^\Gamma}{2\sqrt{\Gamma}} \text{ and } \tau_B = \frac{3\eta V_H}{k_B T} \quad (11)$$

where $\tau_0$ is the average relaxation time in response to a thermal fluctuation, $\eta$ the viscosity of the medium, $k_B$ the Boltzmann constant, $T$ the absolute temperature, and $\Gamma = K V_M/(k_B T)$ with $K$ and $V_M$ being the anisotropy constant of MNPs and the magnetic volume given by $V_M = \pi D^3/6$ for MNPs of diameter $D$, respectively. $V_H$ in equation (11) was taken as the hydrodynamic volume of MNPs that is larger than $V_M$. As a model for $V_H$ it was assumed that $V_H = (1+2\delta/D)^3 V_M$, where $\delta$ is the thickness of a surfactant layer [15]. Because the actual equilibrium susceptibility $\chi_0$ is dependent on the magnetic field, $\chi_0$ was assumed to be the chord susceptibility corresponding to the Langevin equation, given by [16]

$$\chi_0 = \chi_i \frac{3}{\xi}\left[\coth(\xi) - \frac{1}{\xi}\right] \quad (12)$$

where $\chi_i = \mu_0 \phi M_d^2 V_M/(3k_B T)$ and $\xi = \mu_0 M_d H(t) V_M/(k_B T)$. $M_d$ and $\phi$ are the domain magnetization of a suspended particle and the volume fraction of MNPs, respectively. It should be noted that $\chi_0$ and $\xi$ are dependent of time $t$ and that when $\xi$ is zero, $\chi_0$ becomes equal to $\chi_i$. Substituting equations (7) and (8) into equation (6) yields

$$M(t) = H_{ac}[\chi'\cos(2\pi f_{ac} t) + \chi''\sin(2\pi f_{ac} t)] \quad (13)$$

### 2.2.3. Particle size distribution

Since not all particles in a certain volume have the same diameter $D$, the magnetization of MNPs with diameter $D$ at time $t$ ($M_D(t)$) should be averaged based on the particle size distribution as

$$\langle M(t) \rangle = \int_0^\infty M_D(t) \rho(D) dD \quad (14)$$

where $\rho(D)$ denotes the probability density function of the particle size distribution. The result of a natural growth process during particle synthesis does not yield particles with a single diameter $D$, but with a polydisperse particle size distribution [17]. A reasonable and commonly used approach for modeling is the log-normal distribution [17]. The probability density function of the particle size distribution $\rho(D)$ is given by

$$\rho(D) = \frac{1}{\sqrt{2\pi}\sigma D} \exp\left\{-\frac{1}{2}\left[\frac{\ln(D)-\mu}{\sigma}\right]\right\} \quad (15)$$

where $\mu$ and $\sigma$ are the mean and SD of the log-normal distribution, respectively, and are given by

$$\mu = \ln[E(D)] - \frac{1}{2}\ln\left[\frac{\text{Var}(D)}{E^2(D)} + 1\right] \quad (16)$$

and

$$\sigma = \sqrt{\ln\left[\frac{\text{Var}(D)}{E^2(D)} + 1\right]} \quad (17)$$

respectively. $E(D)$ and $\sqrt{\text{Var}(D)}$ denote the mean and SD of $D$, respectively.

### 2.2.4. Calculation of the 3rd and 5th harmonic signals

To investigate the relationship between the harmonic signal and viscosity, we calculated the 3rd and 5th harmonic signals for viscosities ranging from 0.001 to 1.0 kg/m/s. In this study, we considered magnetite ($Fe_3O_4$) as MNPs (SPIONs). Thus, $M_d$, $K$, and $\delta$ were taken as 446 kA/m [18], 9 kJ/m$^3$ [18], and 2 nm [7], [15], respectively. The frequency ($f_{ac}$) and peak-to-peak strength of the AMF ($2 \times H_{ac}$) were fixed at 400 Hz and 20 mT, respectively, in all simulation studies. Temperature was assumed to be room temperature (27 ℃) and 37 ℃. Regarding the particle size of MNPs, we took $\sigma$ in equation (15) and $E(D)$ in equation (16) as 0.282 and 17.1 nm, respectively [7]. It should be noted that Var($D$) can be obtained from $\sigma$ and $E(D)$ using equation (17), and $\mu$ can be obtained from $E(D)$ and Var($D$) using equation (16).

In the simulations, we calculated the $\langle M(t)\rangle$ values at a sampling duration of 100 μs for a total sampling time of 1 s using equation (14), and then calculated the 3rd and 5th harmonic signals from $\langle M(t)\rangle$ using equation (5). In this study, the integration in equation (14) was performed by use of the trapezoidal rule [19].

## 3. Results

Figure 3 shows the normalized 3rd (a) and 5th harmonic signals (b) as a function of time in the case when blood coagulation was induced by adding 0.5 M CaCl$_2$ solution to the whole blood of sheep. The case when CaCl$_2$ solution was not added is also shown for comparison. The closed circles, open circles, and closed squares show groups in which CaCl$_2$ solution was not added to whole blood and kept at room temperature (27 °C), CaCl$_2$ was added and kept at room temperature (27 °C), and CaCl$_2$ was added and kept at 37 °C, respectively. As shown in figure 3, when CaCl$_2$ solution was added, the 3rd harmonic signal decreased with time, and this decrease was enhanced with increasing temperature. Although the 5th harmonic signal also showed similar results, the variation in the signal was larger than that of the 3rd harmonic signal.

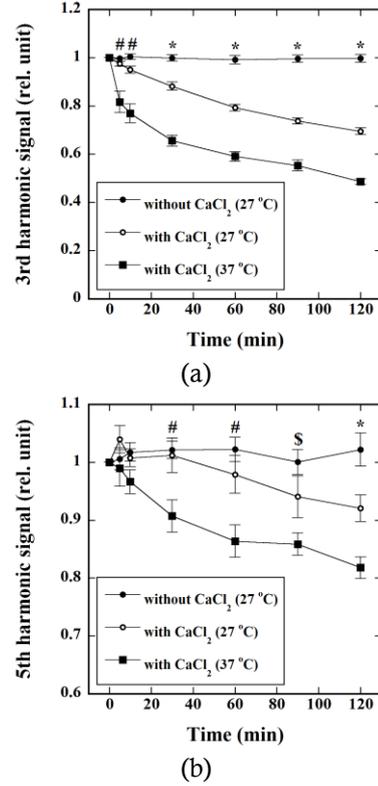

(a)

(b)

**Figure 3:** Time courses of the 3rd (a) and 5th harmonic signals (b) as a function of time in groups in which CaCl$_2$ solution was not added to whole sheep blood and kept at room temperature (27 °C) (closed circles), CaCl$_2$ was added and kept at room temperature (27 °C) (open circles), or CaCl$_2$ was added and kept at 37 °C (closed squares). The signal intensities were normalized by those immediately before adding CaCl$_2$ solution. The peak-to-peak strength and frequency of AMF were taken as 20 mT and 400 Hz, respectively. Data are represented by mean±standard error (SE) (n=5). The symbol * indicates significant differences for all combinations of groups, whereas # indicates significant differences between groups shown by closed circles and open circles and between groups shown by closed circles and closed squares, while $ indicates a significant difference only between groups shown by closed circles and closed squares.

Figure 4 shows column plots of the $k$ (a) and $S_{inf}$ values (b) obtained by fitting the time course of the normalized 3rd harmonic signal after adding CaCl$_2$ solution (figure 3(a)) to equation (1) for groups in which the temperature was kept at room temperature (27 °C) and 37 °C. As shown in figure 4, there were significant differences in both the k and $S_{inf}$ values between the two groups, indicating that the temperature significantly affected blood coagulation.

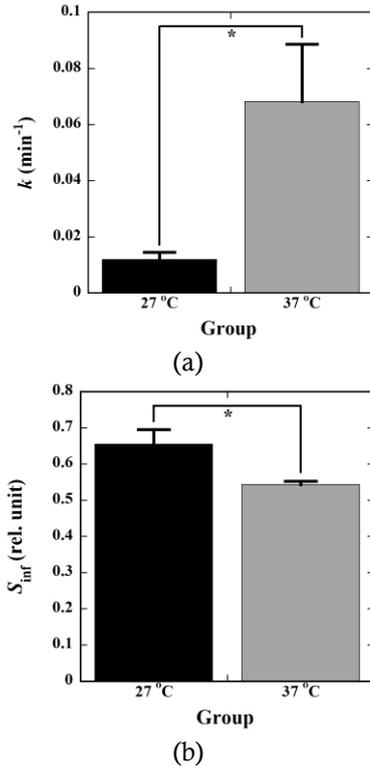

**Figure 4:** Column plots of the coagulation rate ($k$) (a) and the normalized signal intensity at infinite time ($S_{inf}$) (b) calculated by fitting the time course of the normalized 3rd harmonic signal (figure 3(a)) to equation (1) for groups with CaCl$_2$ solution added and kept at room temperature (27 °C) and 37 °C. * $P<0.05$.

Figure 5 shows the normalized 3rd (a) and 5th harmonic signals (b) as a function of time after adding CaCl$_2$ solution to the blood of sheep with various hematocrits. The closed circles, open circles, and closed squares show groups in which the hematocrit was taken as 20%, 40%, and 60%, respectively. As shown in figure 5(a), there were significant differences in all combinations of groups at 5, 10, and 30 min after adding CaCl$_2$ solution. Although there were significant differences between groups with hematocrits of 20% and 60% and between 40% and 60% at 60 min or more after adding CaCl$_2$ solution, a significant difference between 20% and 40% disappeared. As for the 5th harmonic signal (figure 5(b)), the signal at a hematocrit of 60% was always significantly greater than those at 20% and 40%, although there were no significant differences between groups with hematocrits of 20% and 40%.

Figure 6 shows column plots of the $k$ (a) and $S_{inf}$ values (b) obtained by fitting the time course of the normalized 3rd harmonic signal after adding CaCl$_2$ solution (figure 5(a)) to equation (1) for groups in which the hematocrit was taken as 20%, 40%, and 60%. As shown in figure 6(a), there were significant differences in $k$ for all combinations of groups. Although there was a tendency for $S_{inf}$ to increase with increasing hematocrit, it did not reach statistical significance (figure 6(b)).

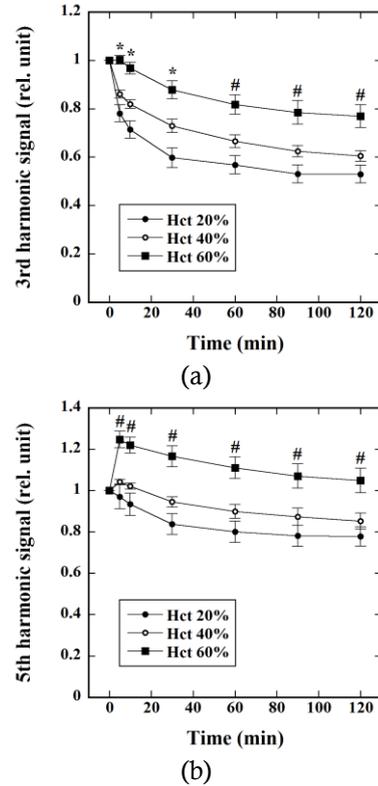

**Figure 5:** Time courses of the 3rd (a) and 5th harmonic signals (b) as a function of time after adding CaCl$_2$ solution in groups in which the hematocrit (Hct) was taken as 20% (closed circles), 40% (open circles), and 60% (closed squares). All measurements were performed at room temperature (27 °C), and the signal intensities were normalized by those immediately before adding CaCl$_2$ solution. The peak-to-peak strength and frequency of AMF were taken as 20 mT and 400 Hz, respectively. Data are represented by mean±SE (n=6). The symbol * indicates significant differences for all combinations of groups, whereas # indicates significant differences between groups shown by closed circles and open circles and between groups shown by closed circles and closed squares.

Figure 7 shows the normalized 3rd (a) and 5th harmonic signals (b) as a function of time after the completion of blood sampling in groups with and without addition of heparin sodium. The open and closed circles show groups in which heparin sodium was or was not added to the whole blood sampled from mice, respectively. As shown in figure 7, when heparin sodium was added, both the 3rd and 5th harmonic signals increased slowly with time and tended to saturate. When heparin sodium was not added, although a temporary increase was observed in the 5th harmonic signal at 10 min after blood sampling, both the 3rd and 5th harmonic signals decreased slowly with time. There were significant differences in the 3rd harmonic signal between groups with and without addition of heparin sodium at 10 min or more after the completion of blood sampling, *i.e.*, at 5 min or more after adding heparin sodium, while significant differences were observed in the 5th harmonic signal between them at 30 min or more after blood sampling, *i.e.*, at 25 min or more after adding heparin sodium.

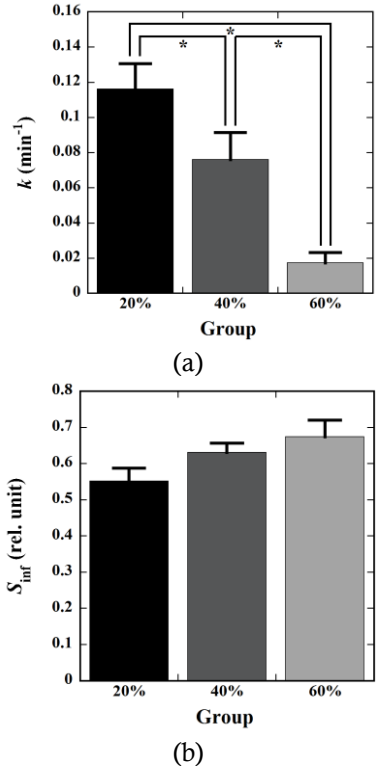

**Figure 6:** Column plots of the $k$ (a) and $S_{inf}$ values (b) calculated by fitting the time course of the normalized 3rd harmonic signal (figure 5(a)) to equation (1) for groups with hematocrits of 20%, 40%, and 60%. * $P<0.05$.

Figure 8 shows the simulation results of the relationship between the 3rd or 5th harmonic signal and viscosity. As shown in figure 8, the 3rd and 5th harmonic signals increased slowly with increasing viscosity and had peaks at approximately 0.015 and 0.025 kg/m/s, respectively. After these peaks, they decreased monotonically with increasing viscosity. Although both the 3rd and 5th harmonic signals decreased slightly with increasing temperature from 27 ℃ to 37 ℃, their dependencies on viscosity mentioned above did not largely differ depending on temperature.

## 4. Discussion

In this study, we presented a method for measuring blood coagulation and its inhibition using SPIONs and AMF. To perform experimental studies, we made an apparatus in which the AMF was generated using a solenoid coil and the 3rd and 5th harmonic signals generated by SPIONs were detected by a gradiometer coil (figure 1). We also performed simulation studies to confirm the rationale of our method. In the simulation studies, it was assumed that the magnetization and particle size distribution of SPIONs obey the Langevin theory of paramagnetism [16] and the log-normal distribution [17], respectively. Our experimental results (figures 3, 5, and 7) demonstrated that our method can detect the process of blood coagulation and its inhibition. Furthermore, we also presented a method for quantifying the coagulation rate from the time course of the 3rd harmonic signal during coagulation process by use of equation (1) (figures 4 and 6). To the best of our knowledge, this is the first report to show that the process of blood coagulation and its inhibition can be detected using SPIONs and AMF and that the coagulation rate can be quantified from the time course of the 3rd harmonic signal during coagulation.

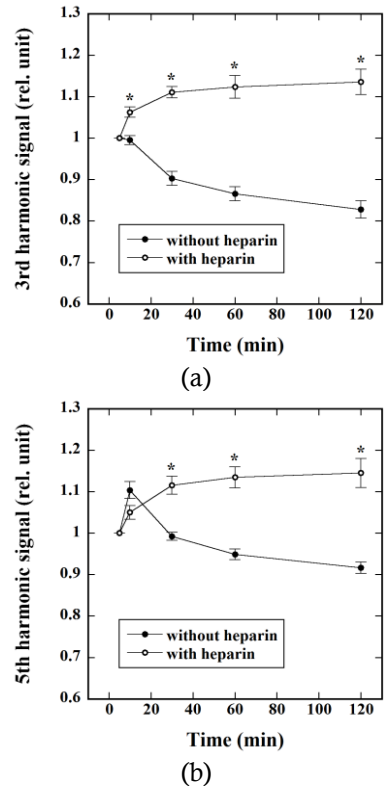

**Figure 7:** Time courses of the 3rd (a) and 5th harmonic signals (b) as a function of time after the completion of blood sampling in groups in which heparin sodium was (open circles) or was not added (closed circles) to the whole blood sampled from mice at 5 min after the completion of blood sampling. All measurements were performed at room temperature (27 ℃), and the signal intensities were normalized by those immediately before adding heparin sodium. The peak-to-peak strength and frequency of AMF were taken as 20 mT and 400 Hz, respectively. Data are represented by mean±SE (n=7). The symbol * indicates a significant difference between the two groups.

Wolberg *et al* [20] reported that blood coagulation is influenced by temperature clinically. They also suggested that bleeding observed at mildly reduced temperatures (33-37 ℃) results primarily from a platelet adhesion defect, and not reduced enzyme activity or platelet activation [20]. However, at temperatures below 33 ℃, both reduced platelet function and enzyme activity likely contribute to the coagulopathy [20]. As shown in figures 3 and 4, our results also demonstrated that blood coagulation is greatly affected by temperature. As shown in figure 4(a), the $k$ value significantly increased with increasing temperature, implying that the coagulation time decreases with increasing temperature. These results are consistent with those reported by Lei et al [21] in which the coagulation time was measured using the electrical impedance method. As shown in the simulation results (figure 8), although both the 3rd and 5th harmonic signals decrease slightly with increasing temperature, the dependency on viscosity does not greatly differ depending on temperature in both the 3rd and 5th harmonic signals. Therefore, the results shown in figures 3 and 4 appear to be mainly due to the change in platelet function and/or enzyme activity in blood depending on temperature [20] rather than the temperature-dependent change of the harmonic signals themselves.

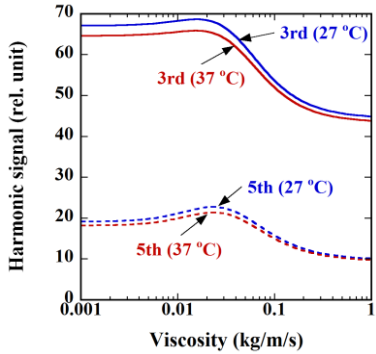

**Figure 8:** Simulation results of the relationship between the 3rd (solid lines) or 5th harmonic signal (dashed lines) and viscosity for temperatures of 27 °C and 37 °C. In these cases, SPIONs were assumed to consist of magnetite ($Fe_3O_4$), and $\sigma$ in equation (15) and $E(D)$ in equation (16) were taken as 0.282 and 17.1 nm, respectively. The peak-to-peak strength and frequency of AMF were assumed to be 20 mT and 400 Hz, respectively.

The fibrinogen in plasma plays an important role in the blood coagulation process. Hematocrit represents the volume percentage of red blood cells in blood. For the blood sample at a low hematocrit, there could be more fibrinogen able to be converted into fibrin fibers than that of blood at a high hematocrit. The mechanical properties of blood clots depend mainly on the formation of fibrin fibers from fibrinogen in the whole blood. The amount of fibrinogen is higher at lower hematocrits, and the viscosity will also be higher [22]. The viscosity of blood clots tends to vary in inverse proportion to the increase of hematocrit [22, 23]. As shown in figure 5, both the 3rd and 5th harmonic signals decreased with decreasing hematocrit. These results appear to support the above concept. Furthermore, as shown in figure 6(a), the $k$ value significantly decreased with increasing hematocrit, implying that the coagulation time increases with increasing hematocrit. Lei et al [21] reported that there was a linear proportional relationship between the blood coagulation time and the hematocrit, based on ultrasonic measurements, which is consistent with our results (figure 6(a)).

Collet et al [24] developed a method for measuring the elastic moduli of individual fibrin fibers in fibrin clots using optical tweezers for trapping beads attached to the fibers that functioned as handles to flex or stretch a fiber. The time courses of the 3rd harmonic signal after adding $CaCl_2$ solution obtained in this study (figures 3(a) and 5(a)) were similar to those of changes in variance in position of a fibrin in blood clots determined from measurement of bead fluctuations as a result of Brownian motion, whose magnitude is inversely related to the stiffness of the fiber [24]. Thus, the decrease in the 3rd and 5th harmonic signals during the blood coagulation process (figures 3 and 5) appears to be due to the fact that the Brownian motion of the MNPs tangled in a meshwork of fibrin fibers is restricted.

It is well known that the out-of-phase component of $\chi$ ($\chi''$) peaks when $2\pi f_{ac}\tau = 1$ [9]. A peak in the harmonic signal occurs near the peak in $\chi''$ where the distortion of the particle magnetization is the greatest. After this peak, the harmonic signal monotonically decreases because the ability of the AMF to align the MNPs is more restricted with increasing viscosity [9]. As shown in our simulation results (figure 8), the 3rd and 5th harmonic signals increased slowly with increasing viscosity and had peaks at approximately 0.015 and 0.025 kg/m/s, respectively. After these peaks, they decreased monotonically with increasing viscosity. These results suggest that if the viscosity of blood is within the range in which the harmonic signals change monotonically, it is possible to detect the process of blood coagulation and anticoagulation using these harmonic signals.

The viscosity of blood was shown to be dependent upon the hematocrit and the velocity distribution or shear rate during flow [27]. The viscosity of whole blood ranged from approximately 0.005 to 0.01 kg/m/s at a normal hematocrit [27]. Although the measured values of viscosity of blood clots vary with the method used, Huang et al [28] reported that the viscosity of the porcine blood clots with hematocrits ranging from 3% to 40% were 0.29 to 0.42 kg/m/s when measured by shear-wave dispersion ultrasound vibrometry. It appears from our simulation results (figure 8) that the above values of viscosity are approximately within the range in which the 3rd and 5th harmonic signals monotonically decrease with increasing viscosity, suggesting the feasibility of detecting blood coagulation using SPOINs and AMF. As shown in figure 8, however, the 5th harmonic signal has a peak at a larger viscosity (approximately 0.025 kg/m/s) than that of the 3rd harmonic signal (approximately 0.015 kg/m/s), suggesting that the range of viscosity within which the 5th harmonic signal decreases monotonically with increasing viscosity is smaller than that of the 3rd harmonic signal. As shown in figures 3(b), 5(b), and 7(b), there was a tendency for the 5th harmonic signal to increase temporarily during blood coagulation. This may be due to the fact that the 5th harmonic signal has a peak during the blood coagulation process as shown in our simulation results (figure 8). Furthermore, the 5th harmonic signal was smaller than the 3rd harmonic signal by a factor of approximately 3 to 4 (figure 8), possibly causing a larger variation in the 5th harmonic signal than that in the 3rd harmonic signal. This finding was consistent with that obtained by our experimental studies (data not shown). Thus, the use of the 3rd harmonic signal appears to be more reliable than that of the 5th harmonic signal for studying blood coagulation or anticoagulation process.

Heparin inhibits blood coagulation by forming complexes with antithrombin III and increasing its affinity to bind and inactivate thrombin and factors IX, Xa, XI, and XII and the tissue factor - VIIa complex [25]. Although it is well known that heparin inhibits thrombin generation, the effect of heparin on blood viscosity has been controversial [26]. Hitosugi *et al* [25] reported that when heparin was added into the blood samples obtained from healthy male volunteers, the blood viscosity decreased and coagulation time increased in a dose-dependent manner. Our results (figure 7) showed that the 3rd and 5th harmonic signals increased slowly with time when heparin was added. These results may reflect the finding reported by Hitosugi *et al* [25].

As shown in figure 1, our apparatus is simple and appears to be easier to implement than other methods such as ultrasonic [2] and microfluidic approaches [3]. As previously described, a new imaging method called MPI has been introduced [5] that allows imaging of the spatial distribution of MNPs with high sensitivity, high spatial resolution, and high imaging speed. If we integrated our method into this MPI system, it would provide a new imaging strategy for detecting blood clots and evaluating the effects of thrombolytic therapy *in vivo*. Furthermore, if we combined our method with magnetic hyperthermia using SPIONs [29], it would produce a new strategy for integrated diagnosis and therapeutics, *i.e.*, theranostics.

In conclusion, we presented a method for measuring blood coagulation and its inhibition using SPIONs and AMF. Our results suggest that our method is useful for measuring them and for studying their processes.

**ACKNOWLEDGEMENT**

The author is grateful to Dr. Iwasaki of Osaka Prefecture University for his help in measuring the size of MNPs and to Mr. Takata and Dr. Shiratsuchi of Osaka University for their help in measuring the magnetization of MNPs.

This work was supported by a Grant-in-Aid for Scientific Research (Grant Number: 25282131) from the Japan Society for the Promotion of Science (JSPS).